\documentclass[aps,pre,preprint,superscriptaddress,nofootinbib]{revtex4-2}

\usepackage[english]{babel}
\usepackage[T1]{fontenc}
\usepackage[utf8]{inputenc}
\usepackage{amsmath,amssymb,bm}
\usepackage{graphicx}
\usepackage[protrusion=true,expansion=false]{microtype}
\usepackage{xcolor}
\usepackage[colorlinks=true,allcolors=blue]{hyperref}

\makeatletter
\let\polymer@frontmatter@abstract@produce\frontmatter@abstract@produce
\def\frontmatter@abstract@produce{%
  \@booleanfalse\preprintsty@sw
  \polymer@frontmatter@abstract@produce
  \@booleantrue\preprintsty@sw
}
\makeatother

\graphicspath{{figures/}}

\newcommand{\kB}{k_{\mathrm B}}
\newcommand{\frec}{f_{\mathrm{rec}}}
\newcommand{\fje}{f_{\mathrm{JE}}}

\begin{document}

\title{A piston-like polymer stochastic heat engine}

\author{Yi-Jui Chiu}
\affiliation{Max Planck Institute of Molecular Cell Biology and Genetics, 01307 Dresden, Germany}

\author{Cheng-Hung Chang}
\affiliation{Institute of Physics, National Yang Ming Chiao Tung University, Hsinchu 300093, Taiwan}

\begin{abstract}
Colloidal stochastic engines are often regarded as microscopic analogues of macroscopic piston–cylinder heat engines. However, although they share some underlying physical principles, such systems remain far from being direct force generators from a practical perspective.
Motivated by this limitation, the present study introduces a polymer-based stochastic engine that more closely mimics the operation of piston–cylinder engines. In this setup, heat is converted into work through a cyclic process in which a polymer is pulled into and out of a narrow channel under varying temperatures.
The work performed by the engine can be directly obtained from the cyclic trajectory in the force–position diagram, analogous to the pressure–volume diagram in traditional heat engines.
Despite its much higher number of degrees of freedom compared to colloidal engines, the polymer engine nevertheless follows several characteristic features observed in such systems. Numerical results demonstrate consistency with universal low-dissipation bounds for efficiency, recovery of Carnot efficiency under regeneration, and low-dissipation scaling of work and power.
\end{abstract}

\maketitle

\section{Introduction}

A heat engine becomes mechanically useful when a temperature difference drives a coordinate that can transmit work to a load, while at mesoscopic scales the work, heat, and entropy production of a single operation fluctuate strongly enough that this conversion must be characterized trajectory by trajectory within stochastic thermodynamics \cite{sekimoto1998langevin,jarzynski1997nonequilibrium,crooks1999entropy,seifert2012stochastic}.  Optical-trap experiments have brought this regime under quantitative control through Stirling and Carnot protocols applied to individual Brownian particles, including carefully designed adiabatic strokes, finite-time operation, and reservoirs with active, non-Gaussian, or memory-bearing fluctuations \cite{blickle2012realization,martinez2015adiabatic,martinez2016brownian,martinez2017colloidal,krishnamurthy2016micrometre,roy2021tuning,chang2023stochastic,krishnamurthy2023overcoming,ginot2025energy}.

The first practical limitation concerns the form in which these engines deliver work, since a driven system with potential $U(x,\lambda)$ exchanges the parameter work $W_{\mathrm{on}}=\int(\partial U/\partial\lambda)\,\mathrm{d}\lambda$, and in the standard harmonic implementation $U(x,k)=kx^2/2$ the controlled parameter is the trap stiffness $k$ while its conjugate observable is $x^2/2$, so compression and expansion appear as changes in the variance of a particle around a nearly stationary mean position \cite{sekimoto1998langevin,blickle2012realization,martinez2016brownian,martinez2017colloidal}.  This construction gives a precise thermodynamic work measurement and a powerful test of finite-time theory, while transmission to another mesoscopic component requires an added transduction stage that converts stiffness modulation into directed displacement, an engineering requirement that has motivated complementary schemes based on translated colloids, cavitation, and orientational dynamics \cite{quintosu2014microscopic,ginot2025energy,dutta2026ellipsoid}.

Autonomous heat engines address the energetic and architectural cost of an externally scheduled protocol by encoding rectification in time-independent couplings, and they can generate steady torque, directed current, rotor acceleration, or self-sustained oscillation from continuously applied thermal or effective stochastic reservoirs \cite{filliger2007brownian,serragarcia2016mechanical,roulet2017autonomous,lin2022stochastic}.  A mechanical resonator implementation has already produced an autonomous force--displacement cycle, although the reported realization was a macroscopic tabletop system driven by an artificial stochastic reservoir, with microscopic operation examined numerically \cite{serragarcia2016mechanical}.  The accompanying limitation for a mesoscopic piston-like device is one of cycle-level control, since autonomous steady-state network engines generate trajectory-resolved cycles that can fluctuate in duration and direction and can compete with additional stochastic cycles \cite{mayrhofer2021stochastic}, while autonomous rotors and self-oscillators tie their cycle geometry to the coupled dynamics of the working substance and load \cite{serragarcia2016mechanical,roulet2017autonomous}.  These architectures are well suited to continuous rotation, current generation, and flywheel charging, while a mechanically resolved task built around a prescribed finite translation calls for independently selected stroke endpoints, branch temperatures, and force levels, so the corresponding cycle area and the work delivered during each branch remain directly accessible.

The externally driven colloidal and autonomous approaches therefore expose complementary parts of the same design problem, with the former retaining a programmable thermodynamic cycle whose output is expressed as parameter work and the latter producing self-organized motion whose cycle geometry follows the rectifier and load dynamics.  A piston-like mesoscopic engine can bridge these capabilities by retaining a branch-resolved thermal cycle while placing its output on a finite translational coordinate, so that the conjugate force is directly measurable and the enclosed force--position area gives the work delivered in one operation.

A partly confined polymer supplies such a coordinate since passage into a narrow channel removes accessible conformations, raises the confinement free energy, and generates an entropic recoil force directed toward the open region, a phenomenon observed in single-DNA nanofluidic experiments and studied through scaling, simulation, and chemical-potential approaches \cite{turner2002confinement,mannion2006conformational,klushin2008dragging,prinsen2009force,yeh2012entropy,xie2019chemical}.  Once a sufficiently long section of the chain spans the channel, the recoil force approaches a plateau along the insertion coordinate, its magnitude grows in proportion to temperature in the ideal hard-wall regime, and, in a prospective nanofluidic implementation, a calibrated electrophoretic force applied to a terminal segment could oppose the recoil and control the stroke \cite{mannion2006conformational,keyser2006direct,prinsen2009force,xie2019chemical}.

We therefore construct a cyclic engine in which one end of the polymer is drawn into the channel at temperature $T_c$ and released at temperature $T_h$, with $T_h>T_c$, so that the two isothermal branches occur at distinct force levels and enclose a loop in the force--position plane whose signed area gives the cycle work,
\begin{equation}
  W_{\mathrm{cyc}}=\oint f\,\mathrm{d}\ell,
  \label{eq:loop_work}
\end{equation}
where $\ell$ is the polymer length displaced toward the open region and $f$ is the outward force.  In the simulations the driven coordinate $\lambda=x_0$ increases into the channel, so $\mathrm{d}\ell=-\mathrm{d}\lambda$, $f=|F_\lambda|$, and $W_{\mathrm{cyc}}=-\oint F_\lambda\,\mathrm{d}\lambda=W_h-W_c$.  The resulting coordinate has a finite range and remains available for coupling to another mesoscopic element, giving the engine the operational feature that motivates the piston-like designation.

The calculations below first establish that the simulated recoil force follows the confinement theory and remains essentially constant over each isothermal stroke, then characterize finite-time work, power, and efficiency through the inverse-time dissipation law developed for low-dissipation engines \cite{esposito2010efficiency,calvohernandez2015time,ma2020experimental,abiuso2020optimal}, and finally compare the polymer engine with efficiency-at-maximum-power bounds and arbitrary-power constraints \cite{curzon1975efficiency,schmiedl2008efficiency,holubec2016maximum,ma2018universal}.  These tests place a many-degree-of-freedom working substance within the finite-time structure previously established for simpler stochastic engines and show how that structure survives when the output is expressed along a translational pulling coordinate.

\section{Results}

\subsection{Construction and quasistatic cycle}

\begin{figure}[t]
  \centering
  \includegraphics[width=0.92\linewidth]{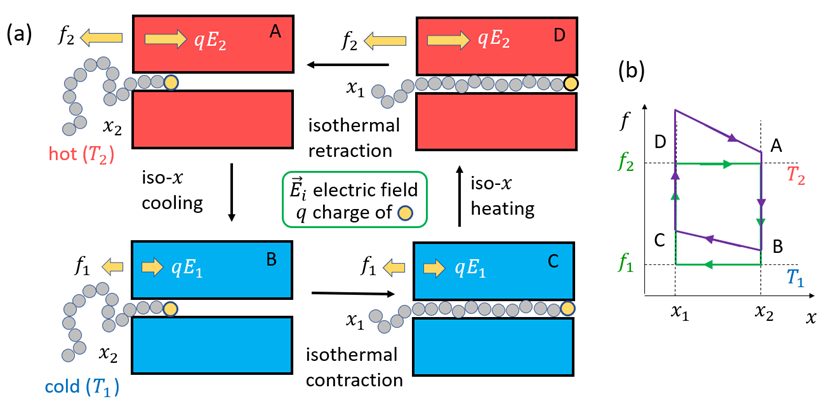}
  \caption{\textbf{Construction and quasistatic cycle of the polymer engine.} \textbf{a}, A terminal bead controls the length of polymer outside the channel.  The labels in the schematic map onto the thermodynamic notation through $T_1=T_c$, $T_2=T_h$, $f_1=f_c$, $f_2=f_h$, $E_1=E_c$, and $E_2=E_h$, while the schematic position labels $x_1$ and $x_2$ correspond to $\ell_1$ and $\ell_2$.  The cycle comprises iso-position cooling $A\to B$, cold isothermal contraction $B\to C$, iso-position heating $C\to D$, and hot isothermal retraction $D\to A$.  The simulations prescribe the terminal coordinate directly.  In the proposed electrophoretic implementation, the calibrated forces $qE_c$ and $qE_h$ would act into the channel, while the entropic recoil forces $f_c$ and $f_h$ act toward the open region, with $q$ denoting the calibrated effective electrophoretic charge.  \textbf{b}, Cycle in the force--position plane.  The green rectangle gives the ideal hard-wall limit with a constant entropy gradient, while the purple trajectory illustrates a generic deformation when energetic polymer--channel interactions contribute to the confinement free energy.}
  \label{fig:engine}
\end{figure}

The polymer engine consists of a bead--spring chain with one section confined in a narrow channel and one terminal bead held at an externally prescribed coordinate, as shown in Fig.~\ref{fig:engine}.  Confinement lowers the configurational entropy and produces an outward recoil force, and the simulations drive the terminal coordinate directly, while a calibrated electrophoretic force applied to that bead offers a prospective experimental means of controlling the amount of polymer that remains outside \cite{keyser2006direct,yeh2012entropy}.  The four states $A$ through $D$ form an iso-position cooling stroke $A\to B$, a cold isothermal contraction $B\to C$, an iso-position heating stroke $C\to D$, and a hot isothermal retraction $D\to A$, with the terminal coordinate traversing the same stroke length $L=\ell_2-\ell_1$ during contraction and retraction.

For a purely entropic confinement free energy $F(\ell,T,d)$, the outward force is
\begin{equation}
  \frec(T,d)
  =-\left(\frac{\partial F}{\partial \ell}\right)_{T,d}
  =T\left(\frac{\partial S}{\partial \ell}\right)_{T,d}.
  \label{eq:entropic_force_general}
\end{equation}
The entropy gradient reaches a constant plateau once the confined and unconfined polymer sections are long enough to suppress end corrections, and the two-dimensional hard-wall model then gives \cite{xie2019chemical}
\begin{equation}
  \frec(T,d)
  =\frac{\kB T}{l_0}
  \ln\!\left(\frac{2\pi l_0}{d}\right),
  \label{eq:entropic_force}
\end{equation}
where $l_0$ is the bead spacing and $d$ is the free lateral width of the channel.  We write $f_c=\frec(T_c,d)$ and $f_h=\frec(T_h,d)$, so the quasistatic input and output work magnitudes are
\begin{equation}
  W_c^{\mathrm{qs}}=f_cL,
  \qquad
  W_h^{\mathrm{qs}}=f_hL,
  \label{eq:qs_branch_work}
\end{equation}
and the work delivered by one cycle becomes
\begin{equation}
  W_{\mathrm{cyc}}^{\mathrm{qs}}
  =W_h^{\mathrm{qs}}-W_c^{\mathrm{qs}}
  =(f_h-f_c)L.
  \label{eq:qs_cycle_work}
\end{equation}

The iso-position strokes exchange heat while producing zero mechanical work.  An ideal regenerator stores the heat released during $A\to B$ and returns it during $C\to D$, which leaves the hot isothermal retraction as the external heat input, $Q_h^{(r)}=f_hL$, and gives
\begin{equation}
  \eta_{\mathrm{qs}}
  =\frac{W_{\mathrm{cyc}}^{\mathrm{qs}}}{Q_h^{(r)}}
  =1-\frac{f_c}{f_h}
  =1-\frac{T_c}{T_h}
  \equiv \eta_C.
  \label{eq:regenerated_carnot}
\end{equation}
All efficiencies reported below use this regenerated-cycle convention, which isolates the two isothermal work-producing branches simulated here.

\subsection{Thermodynamic validation of the isothermal branches}

\begin{figure}[t]
  \centering
  \begin{minipage}[t]{0.485\linewidth}
    \textbf{a}\par
    \includegraphics[width=\linewidth]{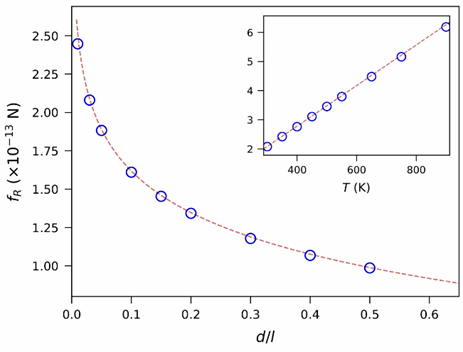}
  \end{minipage}\hfill
  \begin{minipage}[t]{0.485\linewidth}
    \textbf{b}\par
    \includegraphics[width=\linewidth]{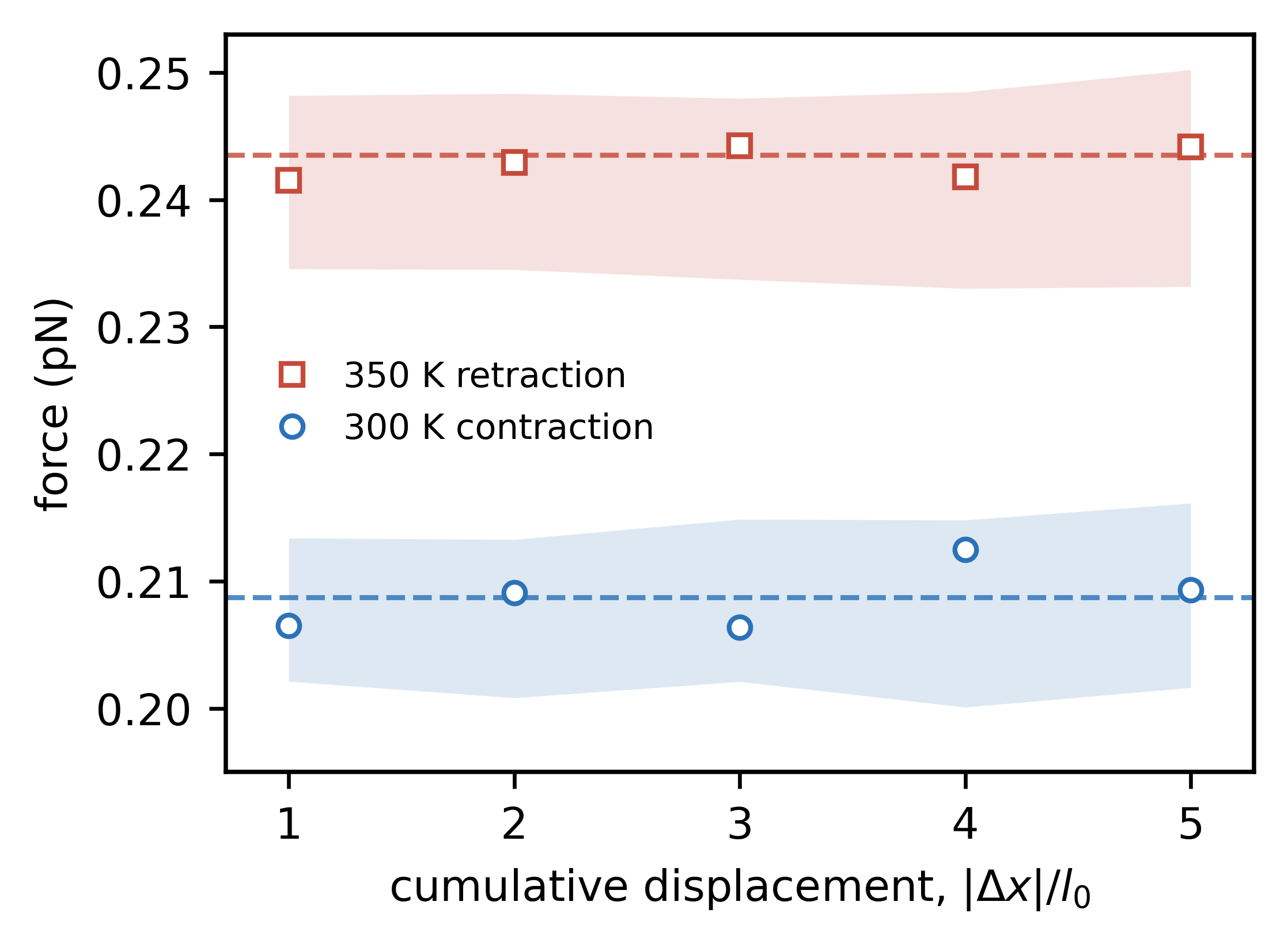}
  \end{minipage}
  \caption{\textbf{Thermodynamic validation of the isothermal branches.} \textbf{a}, Entropic recoil force $f_R\equiv f_{\mathrm{rec}}$ as a function of normalized channel width $d/l$, with $l\equiv l_0$, at $T=300\,\mathrm{K}$.  Open circles show the Jarzynski estimates obtained from $200$ work samples and the dashed curve gives Eq.~\eqref{eq:entropic_force}.  The inset shows the temperature dependence at $d/l=0.03$.  \textbf{b}, Positive force magnitude accumulated over successive one-bead-length intervals during contraction at $300\,\mathrm{K}$ and retraction at $350\,\mathrm{K}$ for $d/l_0=0.03$ and $v=5.0\times10^{-9}\,\mathrm{m\,s^{-1}}$.  Blue circles and red squares show work-preserving interval averages from representative contraction and retraction realizations, shaded regions connect the pointwise 16th--84th percentile ranges across $200$ realizations, and dashed horizontal lines give the theoretical recoil forces at the corresponding temperatures.  The abscissa gives the magnitude of cumulative displacement along each branch.  The representative-realization criterion is specified in Methods.}
  \label{fig:validation}
\end{figure}

The force scale of the engine is set by the confinement free-energy change, which we recover from nonequilibrium work samples through the Jarzynski equality \cite{jarzynski1997nonequilibrium,crooks1999entropy},
\begin{equation}
  \Delta F_{\mathrm{JE}}
  =-\kB T\ln\left\langle
  \exp\!\left(-\frac{W}{\kB T}\right)
  \right\rangle,
  \qquad
  \fje=\frac{|\Delta F_{\mathrm{JE}}|}{L}.
  \label{eq:jarzynski_force}
\end{equation}
Figure~\ref{fig:validation}a shows that the Jarzynski estimates recover both the logarithmic channel-width dependence and the linear temperature dependence of Eq.~\eqref{eq:entropic_force}, extending the same free-energy reconstruction used previously for confined polymers \cite{xie2019chemical} to the parameter range of the present engine.

The engine cycle also requires the force to remain stable as the terminal bead moves over its finite stroke.  We divide each branch into five intervals of length $l_0$ and calculate a work-preserving force magnitude $\overline F_j=|\sum_j\mathrm{d}W/\sum_j\mathrm{d}x_0|$ in every interval, as shown in Fig.~\ref{fig:validation}b.  The distributions remain centred near the corresponding theoretical forces throughout contraction at $300\,\mathrm{K}$ and retraction at $350\,\mathrm{K}$.  Fitting the five-point force profile of each realization separately and averaging the $200$ resulting slopes gives $(2.29\pm2.84)\times10^{-4}\,\mathrm{pN}/l_0$ for contraction and $(-0.40\pm1.68)\times10^{-4}\,\mathrm{pN}/l_0$ for retraction, where the uncertainties are standard errors.  The corresponding two-sided $95\%$ confidence intervals are $[-3.32,7.89]\times10^{-4}\,\mathrm{pN}/l_0$ and $[-3.72,2.92]\times10^{-4}\,\mathrm{pN}/l_0$, so both fitted slopes are statistically consistent with zero.  The measured plateau therefore supplies the pair of iso-force isothermal strokes required by the rectangular quasistatic cycle.

\subsection{Finite-time work, power, and efficiency}

Finite-time driving raises the cold contraction work and lowers the hot retraction work, and the leading slow-driving corrections take the low-dissipation form \cite{esposito2010efficiency,calvohernandez2015time,ma2020experimental}
\begin{equation}
  \langle W_h\rangle
  =W_h^{\mathrm{qs}}-\frac{\Sigma_h}{\tau_h},
  \qquad
  \langle W_c\rangle
  =W_c^{\mathrm{qs}}+\frac{\Sigma_c}{\tau_c},
  \label{eq:low_dissipation_branches}
\end{equation}
where $\tau_h$ and $\tau_c$ are the hot and cold branch durations and $\Sigma_h$ and $\Sigma_c$ have units of $\mathrm{J\,s}$.  The mean cycle work, power, and regenerated efficiency are
\begin{equation}
  W_{\mathrm{cyc}}=\langle W_h\rangle-\langle W_c\rangle,
  \qquad
  P_{\mathrm{cyc}}=\frac{W_{\mathrm{cyc}}}{\tau_{\mathrm{cyc}}},
  \qquad
  \eta=\frac{W_{\mathrm{cyc}}}{\langle W_h\rangle},
  \label{eq:performance_definitions}
\end{equation}
with $\tau_{\mathrm{cyc}}=\tau_h+\tau_c$.  Symmetric protocols use $\tau_h=\tau_c\equiv\tau_b=L/v$ and hence $\tau_{\mathrm{cyc}}=2\tau_b$.

\begin{figure}[t]
  \centering
  \includegraphics[width=\linewidth]{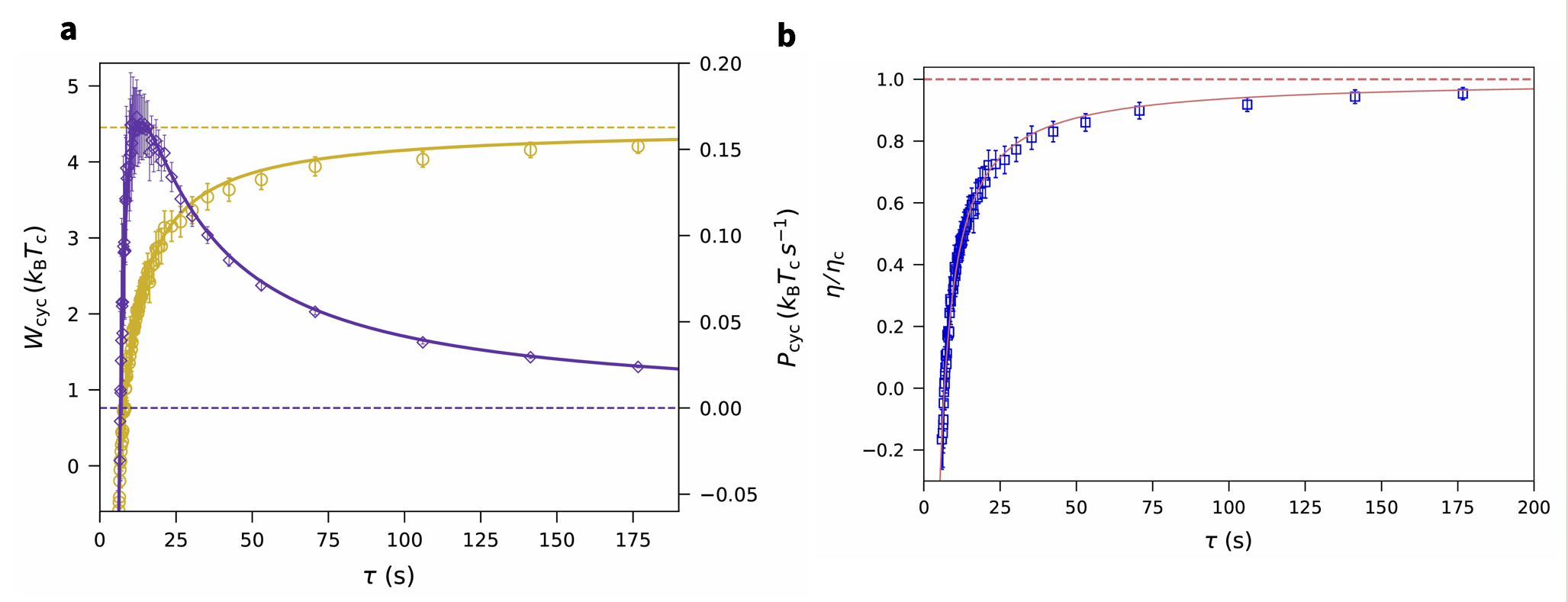}
  \caption{\textbf{Finite-time performance at $T_c=300\,\mathrm{K}$ and $T_h=350\,\mathrm{K}$.} \textbf{a}, Cycle work and power versus total cycle time.  Gold circles show $W_{\mathrm{cyc}}/(\kB T_c)$, the gold solid curve gives the low-dissipation fit, and the gold dashed line gives the quasistatic work.  Purple diamonds show $P_{\mathrm{cyc}}/(\kB T_c\,\mathrm{s}^{-1})$, the purple solid curve is calculated from the branch fits, and the purple dashed line marks zero power.  \textbf{b}, Regenerated efficiency normalized by $\eta_C$.  Blue squares show the simulation values, the coral solid curve follows from the branch fits, and the coral dashed line gives the quasistatic limit.  Error bars give two-sided $90\%$ confidence intervals formed from independent branch ensembles.  The symbol $\tau$ printed on the horizontal axes denotes the total cycle time $\tau_{\mathrm{cyc}}$, and the symbol $\eta_c$ printed in panel \textbf{b} denotes $\eta_C$.}
  \label{fig:finite_time}
\end{figure}

For $T_c=300\,\mathrm{K}$ and $T_h=350\,\mathrm{K}$, the slow-driving data over $\tau_{\mathrm{cyc}}\geq10\,\mathrm{s}$ follow the inverse-duration form of Eq.~\eqref{eq:low_dissipation_branches}, while the theoretical quasistatic cycle work is $W_{\mathrm{cyc}}^{\mathrm{qs}}=4.454\,\kB T_c$.  Figure~\ref{fig:finite_time}a shows the crossover from work-consuming cycles at the fastest protocols to positive work production, followed by a power maximum and the quasistatic work plateau, while Fig.~\ref{fig:finite_time}b shows the corresponding increase of efficiency toward $\eta_C$.  Maximization within the symmetric protocol family gives $\tau_{\mathrm{cyc}}^{*}=13.52\,\mathrm{s}$, $P_{\max}^{\mathrm{sym}}=6.82\times10^{-22}\,\mathrm{J\,s^{-1}}$, and $\eta_{\mathrm{EMP}}^{\mathrm{sym}}=0.0745=0.522\eta_C$.

\subsection{Efficiency bounds and the power--efficiency plane}

\begin{figure}[t]
  \centering
  \includegraphics[width=\linewidth]{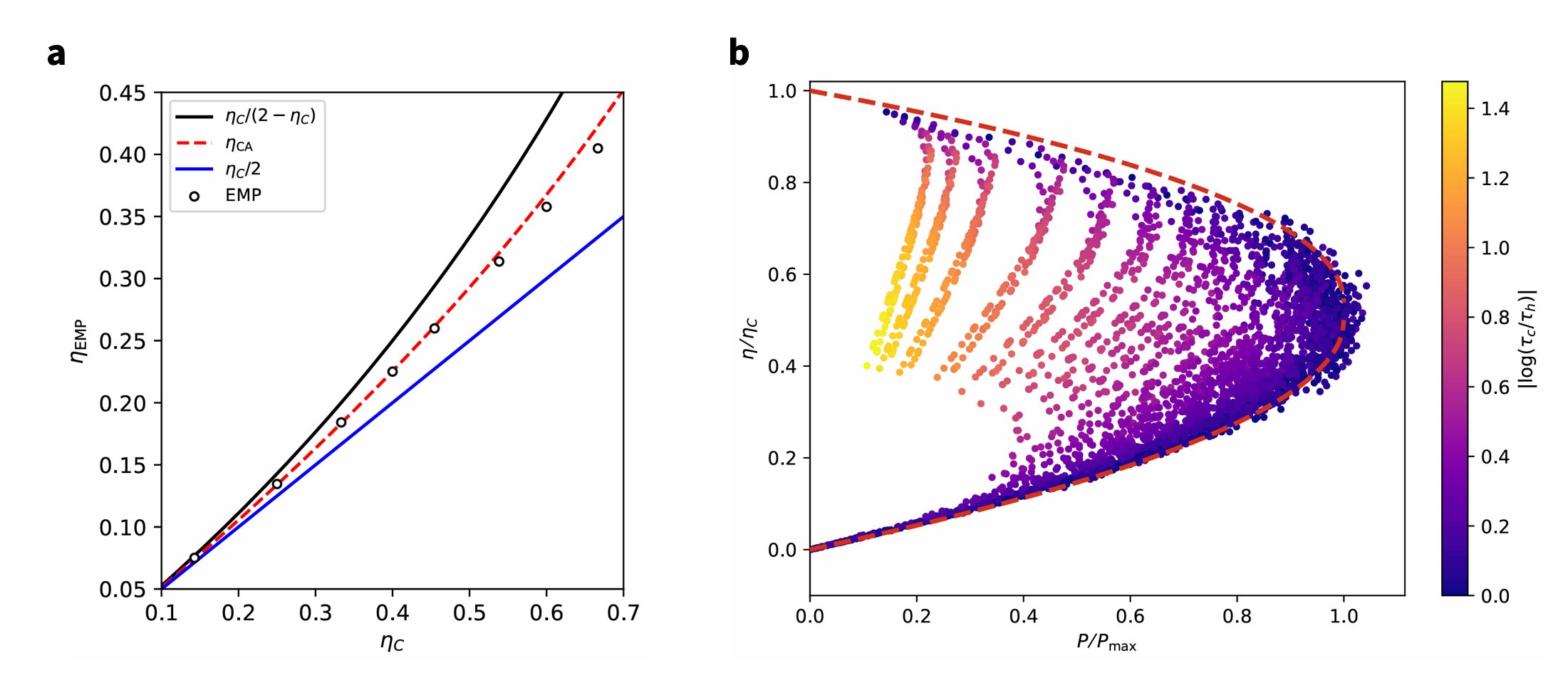}
  \caption{\textbf{Finite-time efficiency bounds.} \textbf{a}, Efficiency at maximum power as a function of Carnot efficiency.  Open circles show EMP values extracted from the fitted equal-duration power curves, the blue and black solid lines give the lower and upper low-dissipation bounds in Eq.~\eqref{eq:emp_bounds}, and the red dashed line gives the Curzon--Ahlborn efficiency.  \textbf{b}, Normalized efficiency--power plane at $T_c=300\,\mathrm{K}$ and $T_h=350\,\mathrm{K}$.  Points show the $3254$ nonnegative-power members of the $58\times58$ Cartesian product of measured hot- and cold-branch summaries.  Colour gives $|\log_{10}(\tau_c/\tau_h)|$, and red dashed curves delimit the low-dissipation region.  The plotted abscissa uses $P_{\max}=P_{\max}^{\mathrm{alloc}}$ from Eq.~\eqref{eq:pmax_asymmetric}.}
  \label{fig:bounds}
\end{figure}

Low-dissipation theory bounds the efficiency at maximum power according to \cite{schmiedl2008efficiency,esposito2010efficiency}
\begin{equation}
  \frac{\eta_C}{2}
  \leq \eta_{\mathrm{EMP}}
  \leq \frac{\eta_C}{2-\eta_C},
  \label{eq:emp_bounds}
\end{equation}
The independent equal-duration temperature scan over $350\leq T_h\leq900\,\mathrm{K}$ yields $0.0754\leq\eta_{\mathrm{EMP}}^{\mathrm{sym}}\leq0.4049$, with all eight points inside the bounds in Fig.~\ref{fig:bounds}a over $0.143\leq\eta_C\leq0.667$ \cite{schmiedl2008efficiency,esposito2010efficiency}.

\begin{samepage}
The complete finite-time performance is characterized by assigning independently sampled durations to the hot and cold branches and normalizing the resulting power by the corresponding allocation optimum
\begin{equation}
  P_{\max}^{\mathrm{alloc}}
  =\frac{\left[\eta_C Q_h^{(r)}\right]^2}
  {4\left(\sqrt{\Sigma_h^{\mathrm{alloc}}}+\sqrt{\Sigma_c^{\mathrm{alloc}}}\right)^2},
  \qquad
  \eta_CQ_h^{(r)}=(f_h-f_c)L.
  \label{eq:pmax_asymmetric}
\end{equation}
\end{samepage}

A single pair of full-range hot- and cold-branch fits defines $\Sigma_h^{\mathrm{alloc}}$ and $\Sigma_c^{\mathrm{alloc}}$ and gives $P_{\max}^{\mathrm{alloc}}=6.880\times10^{-22}\,\mathrm{J\,s^{-1}}$.  We use $P\equiv P_{\mathrm{cyc}}$ in the normalized plot.  Pairing each of the $58$ hot-branch summaries with each of the $58$ cold-branch summaries gives $58^2=3364$ derived allocations at $T_c=300\,\mathrm{K}$ and $T_h=350\,\mathrm{K}$.  The $3254$ nonnegative-power allocations populate the normalized $(P/P_{\max}^{\mathrm{alloc}},\eta/\eta_C)$ plane in Fig.~\ref{fig:bounds}b, while the remaining $110$ negative-power combinations lie outside the displayed engine regime.

\begin{samepage}
The same geometry follows from arbitrary-power optimization and normalized low-dissipation constraints \cite{holubec2016maximum,ma2018universal,zhai2023experimental,zhai2026exact}, while a broader Markovian relation places the result within a more general power--efficiency landscape \cite{shiraishi2016universal}.
\end{samepage}

\section{Discussion}

The central purpose of the polymer engine is to connect stochastic heat conversion with a coordinate that performs work over a finite distance, and partial confinement supplies this connection in a particularly direct form, since every conformational state of the chain contributes to a collective recoil force while the terminal coordinate remains accessible to an external field, a molecular load, or a neighbouring transport element.  The force--position loop therefore carries both the thermodynamic cycle area and the location at which the output can be transmitted, which addresses the practical gap identified for colloidal engines while preserving the trajectory-level energetics that makes those engines theoretically precise \cite{martinez2017colloidal,ginot2025energy}.

The simulations also show that the many internal modes of the polymer preserve a compact finite-time organization over the present operating window, since the branch dissipation is captured by one inverse-time coefficient per isotherm, the resulting maximum-power efficiencies remain between the universal low-dissipation bounds, and the full duration allocation populates the predicted power--efficiency region.  This agreement turns the polymer from a formal replacement for a single trapped coordinate into a test of how universal finite-time structures emerge from an extended working substance with conformational relaxation.

The present model isolates that mechanism through a two-dimensional bead--spring chain, hard reflecting channel walls, overdamped dynamics, prescribed thermal reservoirs, and ideal regeneration during the iso-position strokes, while an experimental realization will add screened electrostatics, hydrodynamic coupling, temperature-switching times, wall chemistry, and the response of a finite load.  Each of these ingredients can be introduced through a measurable modification of the force--position loop, and the plateau force, branch work distributions, and efficiency--power plane established here provide direct observables for determining how the engine changes as those physical effects enter.  Polymer confinement therefore offers a promising path from fluctuating heat conversion to mechanically addressable work, with the finite stroke enabling future studies of particle transport, energy transfer, and load-dependent operation in a single stochastic machine.

\section{Methods}

\subsection{Bead--spring model and driven protocol}

We model the working substance as a two-dimensional polymer of $N=40$ beads indexed by $i=0,\ldots,N-1$, joined by harmonic springs of natural length $l_0=1.06\times10^{-7}\,\mathrm{m}$ and spring constant $k_s=10^{-3}\,\mathrm{N\,m^{-1}}$.  The spring energy is
\begin{equation}
  U_{\mathrm{chain}}
  =\sum_{i=0}^{N-2}\frac{k_s}{2}
  \left(r_{i,i+1}-l_0\right)^2,
  \qquad
  r_{i,i+1}=|\bm r_{i+1}-\bm r_i|.
  \label{eq:spring_energy}
\end{equation}
The terminal bead $i=0$ is driven along the channel axis through the control coordinate $\lambda(t)=x_0(t)$, while beads $i=1,\ldots,N-1$ follow overdamped Langevin dynamics,
\begin{equation}
  \gamma\dot{\bm r}_i
  =-\nabla_iU_{\mathrm{chain}}
  +\sqrt{2\gamma\kB T}\,\bm\xi_i(t),
  \label{eq:langevin}
\end{equation}
with $\gamma=10^{-8}\,\mathrm{kg\,s^{-1}}$ and Gaussian white noise satisfying
\begin{equation}
  \langle\xi_{i\alpha}(t)\rangle=0,
  \qquad
  \langle\xi_{i\alpha}(t)\xi_{j\beta}(t')\rangle
  =\delta_{ij}\delta_{\alpha\beta}\delta(t-t').
  \label{eq:noise}
\end{equation}
The equations are integrated by Euler--Maruyama with $\Delta t=10^{-6}\,\mathrm{s}$.

The channel entrance lies at $x=x_{\mathrm{ch}}$, the channel interior occupies $x\geq x_{\mathrm{ch}}$ and $|y|\leq d/2$, and the main engine calculations use $d=0.03l_0$.  Trial steps that cross a side wall or the solid part of the channel mouth are reflected geometrically until the bead returns to the allowed domain.  The terminal bead retains a fixed transverse coordinate and advances according to
\begin{equation}
  x_0^{(n+1)}=x_0^{(n)}+s\,v\Delta t,
  \qquad
  s=+1\ \text{for contraction},
  \quad
  s=-1\ \text{for retraction}.
  \label{eq:driven_bead}
\end{equation}
The contraction and retraction branches begin at $(x_0-x_{\mathrm{ch}})/l_0=4.5$ and $9.5$, respectively, and traverse this common interval in opposite directions, giving $L=5l_0$.  For each isothermal branch, $200$ initial configurations were generated independently by placing the $40$-bead chain in a straight conformation, holding the driven bead at $y_0=0$ and at the corresponding branch endpoint, and evolving beads $i=1,\ldots,39$ at the bath temperature for $30\,\mathrm{s}$ with $\Delta t=10^{-6}\,\mathrm{s}$, equivalent to $3.0\times10^7$ integration steps.  The contraction and retraction libraries were thermalized separately at $300\,\mathrm{K}$ and $350\,\mathrm{K}$, respectively, and each production trajectory began immediately from its seed-matched configuration.  The same preparation was repeated at the corresponding channel width and bath temperature for the validation scans.

The generalized force conjugate to $\lambda$ is defined by the work convention of stochastic energetics \cite{sekimoto1998langevin,seifert2012stochastic},
\begin{equation}
  F_\lambda
  =\frac{\partial U_{\mathrm{chain}}}{\partial\lambda}
  =k_s(r_{0,1}-l_0)\frac{x_0-x_1}{r_{0,1}},
  \qquad
  \mathrm{d}W=F_\lambda\,\mathrm{d}\lambda.
  \label{eq:conjugate_force}
\end{equation}
Here $W$ is work performed on the polymer, so the cold-branch input magnitude is $W_c=W>0$ and the hot-branch output magnitude is $W_h=-W>0$ for the adopted directions.  The positive force shown in Figs.~\ref{fig:engine} and \ref{fig:validation} is the corresponding outward or pulling-force magnitude.

\subsection{Free-energy and iso-force validation}

For every width or temperature in Fig.~\ref{fig:validation}a, the Jarzynski estimator in Eq.~\eqref{eq:jarzynski_force} uses all $200$ independent work values, and its uncertainty is the standard deviation of $1000$ bootstrap estimates generated with a fixed random seed.  The theoretical comparison uses Eq.~\eqref{eq:entropic_force} with the exact SI value $\kB=1.380649\times10^{-23}\,\mathrm{J\,K^{-1}}$.

The iso-force calculation uses $200$ contraction realizations at $300\,\mathrm{K}$ and $200$ retraction realizations at $350\,\mathrm{K}$, all driven at $v=5.0\times10^{-9}\,\mathrm{m\,s^{-1}}$.  Each trajectory is partitioned into five one-bead-length intervals and its work-preserving force magnitude is evaluated from the summed work and displacement in that interval.  The shaded ranges in Fig.~\ref{fig:validation}b are the pointwise 16th and 84th percentiles of the $200$ values.  Ordinary least squares is applied separately to the five interval-force values of every realization, and the reported slope is the mean of the $200$ trajectory-level slopes, with its standard error calculated as $s/\sqrt{200}$ from their sample standard deviation $s$ and its two-sided $95\%$ confidence interval calculated from Student's $t$ distribution with $199$ degrees of freedom.  Representative realizations are selected reproducibly by ranking each trajectory through its relative root-mean-square deviation from Eq.~\eqref{eq:entropic_force}, retaining the $15$ smallest deviations on each branch, and choosing the pair with the smallest summed deviation under the residual-pattern condition $|\rho|\leq0.30$.  This procedure selects contraction seed $217$ and retraction seed $44$, with $\rho=-0.276$.

\subsection{Finite-time analysis}

The symmetric-cycle data combine independent cold contraction and hot retraction ensembles at $58$ common drive speeds from $0.6\times10^{-8}$ to $18\times10^{-8}\,\mathrm{m\,s^{-1}}$.  Each branch and speed was simulated using $200$ independently thermalized pulling trajectories, and branch standard errors are combined in quadrature.  The low-dissipation coefficients in Eq.~\eqref{eq:low_dissipation_branches} are fitted with the quasistatic work magnitudes fixed by Eq.~\eqref{eq:qs_branch_work}, using the $33$ points satisfying $\tau_{\mathrm{cyc}}\geq10\,\mathrm{s}$.  The curves in Fig.~\ref{fig:finite_time} use these fitted branch functions over the displayed time range, and error bars equal $1.645$ times the propagated standard error, corresponding to two-sided $90\%$ confidence intervals.

For the efficiency-at-maximum-power comparison, the fitting procedure is applied independently at $T_c=300\,\mathrm{K}$ and eight hot temperatures from $350$ to $900\,\mathrm{K}$, with the maximum selected on a $400$-point duration grid along each fitted equal-duration power curve.  For the asymmetric-duration analysis at $T_h=350\,\mathrm{K}$, the $58$ empirical hot-branch summaries and $58$ empirical cold-branch summaries are combined as a Cartesian product, giving $3364$ derived protocol allocations.  Each allocation is evaluated from the corresponding empirical branch means through Eq.~\eqref{eq:performance_definitions}.  The $110$ negative-power combinations lie outside the engine regime displayed in Fig.~\ref{fig:bounds}b.  The allocations carry shared-input correlations, since every branch summary contributes to $58$ pairings, and they therefore do not constitute $3364$ independent full-cycle observations.  The normalization $P_{\max}^{\mathrm{alloc}}$ is calculated from Eq.~\eqref{eq:pmax_asymmetric} using a single pair of full-range hot- and cold-branch fits.

\begin{acknowledgments}
We thank Van Tan Vu for useful discussions. This work was supported by the National Science and Technology Council, Taiwan, under Grant No.~NSTC 113-2112-M-A49-019.
\end{acknowledgments}

\bibliography{references}

\end{document}